\documentclass{aa}
\usepackage{graphicx}
\usepackage{txfonts}

\usepackage{natbib,twoopt}
\usepackage[breaklinks=true]{hyperref} 
\bibpunct{(}{)}{;}{a}{}{,} 
\makeatletter
\newcommandtwoopt{\citeads}[3][][]{\href{http://adsabs.harvard.edu/abs/#3}%
{\def\hyper@linkstart##1##2{}%
\let\hyper@linkend\@empty\citealp[#1][#2]{#3}}}
\newcommandtwoopt{\citepads}[3][][]{\href{http://adsabs.harvard.edu/abs/#3}%
{\def\hyper@linkstart##1##2{}%
\let\hyper@linkend\@empty\citep[#1][#2]{#3}}}
\newcommandtwoopt{\citetads}[3][][]{\href{http://adsabs.harvard.edu/abs/#3}%
{\def\hyper@linkstart##1##2{}%
\let\hyper@linkend\@empty\citet[#1][#2]{#3}}}
\newcommandtwoopt{\citeyearads}[3][][]%
{\href{http://adsabs.harvard.edu/abs/#3}
{\def\hyper@linkstart##1##2{}%
\let\hyper@linkend\@empty\citeyear[#1][#2]{#3}}}
\makeatother

\usepackage{diagbox}
\usepackage{xcolor}

\makeatletter
\renewcommand*\aa@pageof{, page \thepage{} of \pageref*{LastPage}}
\makeatother

\begin{document}

   \title{Modeling the effects of clumpy winds in the high-energy light~curves of $\gamma$-ray binaries}

   \author{E. Kefala
          \inst{}
          \and
          V. Bosch-Ramon\inst{}
          }

   \institute{
   Departament de Física Quàntica i Astrofísica (FQA), Universitat de Barcelona (UB), Martí i Franqués, 1, 08028 Barcelona, Spain \\
Institut de Ciències del Cosmos (ICCUB), Universitat de Barcelona (UB), Martí i Franqués, 1, 08028 Barcelona, Spain \\
Institut d'Estudis Espacials de Catalunya (IEEC), Gran Capità, 2-4, 08034 Barcelona, Spain
\\
           \email{ekefala@icc.ub.edu}\\ \email{vbosch@fqa.ub.edu}
             }

   \date{Received; accepted }


\titlerunning{Modeling the effects of clumpy winds in the high-energy light curves of $\gamma$-ray binaries}

  \abstract
   {High-mass gamma-ray binaries are powerful nonthermal galactic sources, some of them hosting a pulsar whose relativistic wind interacts with a likely inhomogeneous stellar wind. So far, modeling these sources including stellar wind inhomogeneities has been done using either simple analytical approaches or heavy numerical simulations, none of which allow for an exploration of the parameter space that is both reasonably realistic and general.}
   {Applying different semi-analytical tools together, we study the dynamics and high-energy radiation of a pulsar wind colliding with a stellar wind with different degrees of inhomogeneity to assess the related observable effects. 
   }
   {We computed the arrival of clumps to the pulsar wind--stellar wind interaction structure using a Monte Carlo method and a phenomenological clumpy-wind model. The dynamics of the clumps that reach deep into the pulsar wind zone was computed using a semi-analytical approach. This approach allows for the characterization of the evolution of the shocked pulsar wind region in times much shorter than the orbital period. With this three-dimensional information about the emitter, we applied analytical adiabatic and radiative models to compute the variable high-energy emission produced on binary scales.}
   {An inhomogeneous stellar wind induces stochastic hour-timescale variations in the geometry of the two-wind interaction structure on binary scales. Depending on the degree of stellar wind inhomogeneity, 10--100\% level hour-scale variability in the X-rays and gamma rays is predicted, with the largest variations occurring roughly once per orbit. 
   }
   {Our results, based on a comprehensive approach, show that present X-ray and future very-high-energy instrumentation can allow us to trace the impact of a clumpy stellar wind on the shocked pulsar wind emission in a gamma-ray binary.}

   \maketitle
%

\section{Introduction}\label{intro}

High-mass gamma-ray binaries are among the most luminous high-energy sources in the Galaxy. These systems host a massive star and a compact object, and their radiation extends from radio waves to high-energy (HE; 0.1--100~GeV) and very-high-energy (VHE; $>100$~GeV) gamma rays, with the nonthermal emission being dominated by energies $>1$~MeV \citep[see, e.g.,][for a review]{Dubus2013,Paredes2013,Paredes2019}. Observations of these systems also show gamma-ray orbital modulation, indicating that the related emitting region cannot be located too far from the binary \citep[e.g., for LS 5039 and LS~I~+61~303;][respectively]{Aharonian2006,Albert2006}. Approximately ten systems pertaining to this class have been detected so far, but the power origin has not been firmly identified yet in most of them, with the important exceptions of PSR B1259-63 \citep{Johnston1992}, PSR J2032+4127 \citep{Lyne2015}, and probably LS~I~+61~303 \citep{Weng2022}, in which the recent detection of radio pulsations strongly points at the presence of a pulsar (see also \citealt{Yoneda2020} and \citealt{Volkov2021} for the contested detection of X-ray pulses from LS~5039). The uncertainty in the power origin in gamma-ray binaries has fueled a debate between two competing scenarios: the accretion-powered and the pulsar-wind-powered scenarios (e.g., \citealt{Maraschi1981,Paredes2000,Martocchia2005,Dubus2006,Romero2007,Bosch-Ramon2009,Massi2017}; see however, e.g., \citealt{Papitto2012} for a pulsar model with regime transitions). 

In the accretion-powered scenario, particle acceleration is typically thought to occur in the relativistic jets of a microquasar, which are fed by the accretion of stellar material onto the compact object. Jet propagation through the stellar wind medium on binary scales has been numerically studied in the relativistic \citep[e.g.,][]{Perucho2008,Perucho2010,Charlet2022} and nonrelativistic regimes \citep[e.g.,][]{Yoon2015,Yoon2016}. Leptonic and hadronic models have been adopted for the broadband emission from the relativistic jets of high-mass microquasars \citep[e.g.,][]{Romero2008}. The effects of orbital motion on the jet kinematics and the broadband nonthermal emission have also been considered in several (semi-)analytical and numerical works \citep[see, e.g.,][for some recent studies including orbital effects]{Bosch-Ramon2016,Molina2018,Molina2019,Barkov2021b}.

In the standard pulsar-powered scenario, a significant fraction of the pulsar spin-down power is dissipated at the interaction of a relativistic wind from the pulsar and the stellar wind \citep[e.g.,][]{Maraschi1981,Tavani1997}. At the shock front, particles can be efficiently accelerated and emit X-rays via synchrotron and gamma rays via inverse Compton (IC) scattering on the stellar photon field \citep[e.g.,][]{Tavani1994,Kirk1999,Dubus2006,Khangulyan2007}; although, hadronic scenarios have also been proposed \citep[e.g.,][]{Neronov2007}. Relativistic hydrodynamic simulations show that fast reacceleration of the shocked pulsar flow should occur beyond the sonic point \citep[at the outskirts of the binary; see, e.g.,][]{Bogovalov2008}, which would influence the shocked pulsar wind emission by affecting adiabatic losses, the magnetic field, and Doppler boosting \citep[e.g.,][]{Kong2012,Khangulyan2014b,Dubus2015,Molina2020}. The magnetization (and anisotropy) of the pulsar wind can have only a moderate effect on the dynamics and overall structure of the flow \citep{Bogovalov2012}, unless the pulsar wind magnetic field becomes dynamically dominant \citep{Bogovalov2019}. Instability growth and turbulence can impact the structure of the shocks and produce shocked two-wind mixing, and couple with orbit-related Coriolis forces that turn the shocked flows into an unstable spiral structure \citep[e.g.,][]{Bosch-Ramon2011,Bosch-Ramon2012,Lamberts2012,Lamberts2013,Bosch-Ramon2015}. Detailed modeling of the shocked-flow dynamics and radiation in the case of LS 5039, including orbital motion, has been performed by \cite{Molina2020} (semi-analytically), taking the emission at the Coriolis turnover into account, and by \cite{Huber2020a,Huber2020b} (numerically), accounting for the complex fluid dynamics through coupled relativistic hydrodynamics (RHD)--nonthermal particle calculations. The role of eccentricity in the large-scale evolution of the shocked flows has also been explored, using RHD simulation data to estimate the associated nonthermal emission \citep[e.g.,][]{Barkov2018,Barkov2021a}.

Smooth flows are generally considered in the pulsar-wind-powered scenario, but small-scale inhomogeneities can inherently arise from radiation instabilities in the acceleration phase of hot-star winds \citep[e.g.,][]{Lucy1970,Runacres2002,Puls2008}. Large-scale structures may also be present and possibly originate from the circumstellar disks of fast-rotating stars \citep[see, e.g.,][in the context of gamma-ray binaries]{Okazaki2011,Chernyakova2014}, corotating interaction regions between different velocity streams, or magnetically confined regions at the wind base \citep[e.g.,][]{Cranmer1996,Lobel2008}. In nonaccreting systems, the radiative consequences of the presence of clumps were analytically explored, for instance, by \cite{Zdziarski2010} in the context of LS~I~+61~303. A general study, also of analytical nature, of the dynamical and radiative effects of a clumpy stellar wind interacting with a pulsar wind was carried out by \citet{Bosch-Ramon2013a}. Later on, axisymmetric RHD simulations by \cite{Paredes-Fortuny2015} and nonthermal emission computations based on RHD simulation data by \cite{delaCita2017a} were performed to numerically study the effects of the arrival of a single clump in the two-wind interaction region. These works concluded that, depending on its size, a clump can noticeably perturb the interaction region and largely affect the output of the associated nonthermal radiation. For instance, in the mentioned works and in \cite{Chernyakova2014},  it was argued that the collision of a large clump of stellar material with the pulsar wind offers a plausible explanation for the observed gamma-ray flares (e.g., PSR~B1259-63) and quick X-ray variability (e.g., LS~5039, LS~I~+61~303) in high-mass gamma-ray binaries. 
Indeed, short-term variability of high-energy emission seems to be a common feature in gamma-ray binaries \citep[e.g.,][]{Bosch-Ramon2005,Smith2009,Takahashi2009,Rea2011,An2015,Tam2018}.
Studies of the clumpy-wind presence have also been performed, for instance, in the context of massive-star binaries \citep{Pittard2007}, accreting X-ray sources \citep[e.g.,][]{Oskinova2012}, and microquasars \citep[e.g.,][]{Araudo2009,Owocki2009,Perucho2012,delaCita2017b,Lopez2022}.

\begin{figure}
\centering
     \centering\resizebox{1\hsize}{!}{\includegraphics{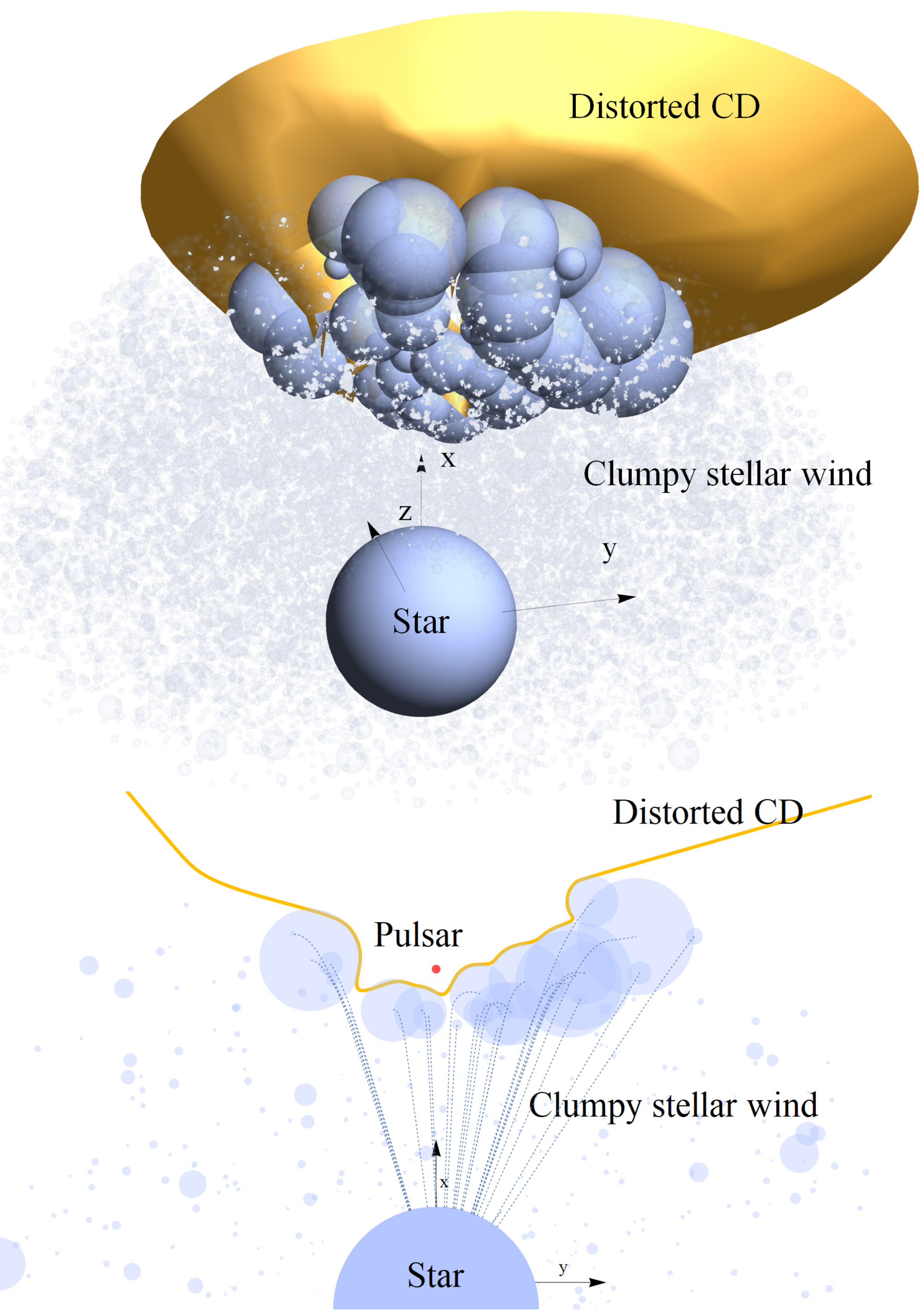}}
     \caption{Three-dimensional schematic (top) and two-dimensional cut (bottom) of the physical scenario comprising a massive star with a clumpy stellar wind and a pulsar. The dashed lines in the bottom panel show the trajectories of clumps that have crossed the smooth-wind contact discontinuity.}
     \label{sketch}
\end{figure}

As explained above, so far only simple analytical treatments or costly numerical calculations have been used to model the emitting flow under a clumpy stellar wind in the pulsar-powered scenario. Here, we present a novel approach to modeling the effects of a clumpy stellar wind on the emitting region in that scenario; this approach aims at balancing realistic modeling and computational efficiency. The unshocked and shocked clump propagation and dynamics are modeled adopting semi-analytical Monte Carlo and hydrodynamic calculations, allowing for a general exploration of the effects of multiple clumps on the geometry evolution of the shocked two-wind region. The associated nonthermal emission and its variability induced by the clumps are analytically modeled treating the emitting particles either in an adiabatic or a radiative regime, with a particle power-law energy distribution. Gamma-ray absorption is taken into account. This method is very fast to implement but still provides the most relevant information; although, formally it can only be applied consistently to the region where the shocked pulsar wind is subsonic. The paper is structured as follows: the physical scenario and its modeling are explained in Sect.~\ref{numa}; the results are presented in Sect.~\ref{res}; and a summary and discussion are provided in Sect.~\ref{disc}.

\section{Physical system and modeling}\label{numa}
\subsection{Physical system}

Although this study applies to high-mass gamma-ray binaries hosting young pulsars in general, we consider a system with orbit and distance similar to those of LS~5039 \citep[e.g.,][]{Casares2005}, a powerful and well-studied source. The adopted components for the binary system are a main-sequence O-type star with luminosity $L_{*}=10^{39}$~erg~s$^{-1}$ and temperature $T=4 \times 10^{4}$~K, and a nonaccreting pulsar with spin-down power set to $P_{\rm sd}=3\times 10^{36}$~erg~s$^{-1}$. The orbital parameters are the system eccentricity $e=0.35$, orbital period $P=3.91$~d, and semimajor axis $\alpha =2.1 \times 10^{12}$~cm. The orientation of the system with respect to the observer is such that superior and inferior conjunction occur at orbital phases $\phi=0.058$ and $\phi=0.716$, respectively. We assume 
a distance of $d_{\rm{obs}}=3$~kpc and 
an inclination with respect to the line of sight of $i=60^{\circ}$ (favored in LS~5039 if it harbors a neutron star; \citealt{Casares2005}). The actual role of $i$ is however secondary for us, as it mostly affects the orbital spectral energy distributions (SEDs) and light curves; for suborbital variability, $i$ mainly influences the normalization of the curves.

The massive star powers a supersonic wind that is driven by line radiation pressure. This wind is assumed to be inhomogeneous, with most of its mass carried by clumps. We adopt a wind mass-loss rate of $\dot{M}=10^{-7}\,$M$_{\odot}$ yr$^{-1}$ and a wind velocity approximated to a constant $u_{\rm w}=2\times 10^{8}$~cm~s$^{-1}$.
The pulsar drives an ultrarelativistic wind that is assumed to be cold, isotropic, and weakly magnetized \citep[see however][and references therein]{Bogovalov2012,Derishev2012,Bogovalov2019,Bosch-Ramon2021}. Typical bulk Lorentz factors are $\Gamma_{\rm{w}}=10^{5}$ \citep{Khangulyan2012,Aharonian2012}, but our results will not strongly depend on this value as the shocked clumps do not accelerate up to relativistic velocities; $\Gamma_{\rm{w}}$ can also affect the energy range of the flow emitting particles \citep[e.g.,][]{Kennel84}, but this has little effect on our conclusions and is neglected. The system parameters are summarized in Table~\ref{table:param}, and a schematic of the system is shown in Fig.~\ref{sketch}.

\subsection{Modeling}

\subsubsection{Smooth-wind contact discontinuity}

The stellar and pulsar winds collide to form a double-bow-shock structure made of shocked wind material. The shape of the contact discontinuity (CD) separating the shocked stellar and pulsar winds can be described, in the smooth-wind case, by a simple approximation of the boundary of equilibrium between the two wind ram pressures \citep{Canto1996}:
\begin{equation}
\label{eq:CD}
\vartheta_{1} =\left[ \frac{15}{2} \left( -1 + \sqrt{1+ \frac{4}{5}\eta (1-\vartheta \rm{cot}\vartheta )}   \right ) \right ]^{1/2},
\end{equation}
where $\vartheta$ and $\vartheta_{1}$ correspond to $\theta$ and $\theta_1$ in Fig.~1 from \cite{Canto1996}. Equation~\ref{eq:CD} gives the shape of a one-dimensional CD cut, but the latter axisymmetry in the smooth-wind case allows us to derive the actual geometry of the two-dimensional CD by rotating this one-dimensional cut around the CD symmetry axis.

\begin{table}
\caption{System parameters}
\label{table:param}
\centering
\begin{tabular}{l l l}   
\hline\hline
Parameter                  &                   & Value                  \\ \hline
Stellar luminosity         & $L_{*}$           & $10^{39}$ erg s$^{-1}$ \\ 
Star temperature           & $T_{*}$           & $4\times 10^{4}$ K \\
Stellar radius             & $R_{*}$           & $10.5\,$R$_\odot$ \\
Stellar mass-loss rate     & $\dot{M}$         & $10^{-7}\,$M$_{\odot}$~yr$^{-1}$ \\
Stellar wind velocity      & $u_{\rm w}$      & $2\times 10^{8}$ cm s$^{-1}$ \\
Pulsar spin-down power     & $P_{\rm{sd}}$     & $3\times 10^{36}$ erg s$^{-1}$ \\
Pulsar wind Lorentz factor & $\Gamma_{\rm{w}}$ & $10^{5}$   \\
Periodicity                & $P$               & $3.91$ d   \\
Eccentricity               & $e$               & $0.35$     \\
Semimajor axis            & $\alpha$          & $2.1 \times 10^{12}$ cm  \\
Inclination                & $i$               & $60^{\circ}$  \\
Distance to observer       & $d_{\rm{obs}}$    & $3$ kpc    \\ \hline
\end{tabular}
\end{table}

The ratio of the pulsar and the stellar wind momentum rates,
\begin{equation}
\eta=(L_{sd}/c)/(\dot{M} u_{\rm w})\,, 
\end{equation}
is $\approx 0.08$ for the selected parameters, a usual value in the literature. Such an $\eta$ value means that the stellar wind momentum rate dominates and the CD bends over the pulsar, but most of the luminosity arriving at the CD still comes from the pulsar wind. The distance of the stagnation point from the pulsar is $R_0=d\eta^{1/2}/(1+\eta^{1/2})$, where $d$ is the orbital separation distance. The calculations are done in the rotating frame associated with the pulsar orbital motion, with $\omega_{\rm orb}$ being the angular velocity of the orbit described above. In this frame, due to the orbit-induced Coriolis force, the CD is slightly tilted with respect to the star--pulsar direction. Here, the tilting of the CD is introduced by rotating the entire structure counter-wise with respect to the orbital motion by an angle 
$\sim d\omega_{\rm orb}/u_{\rm w}$.
Given that most of the clumps are expected to get destroyed in the shocked flow \citep{Pittard2007,Bosch-Ramon2013a,Paredes-Fortuny2015}, the actual CD shape is approximated here as a smooth-case CD plus distortions produced by the largest clumps (see below).

We treat the pulsar wind after the termination shock as subsonic because it strongly simplifies the emitting flow hydrodynamics; although, far enough from the shock the flow can become supersonic again. The region we consider, of size $\lesssim d$, is somewhat larger than the corresponding region in the smooth-wind case \citep{Bogovalov2008}, but flow irregularities induced by the clump presence may justify this choice. Further away from the star, beyond the sonic point, shocked clumps are expected to have already dissolved in the shocked two-wind structure, while the incoming stellar wind tends to be more homogeneous \citep[see, e.g.,][and references therein]{Rubio2022}. These regions however would present flow reacceleration, local instabilities, wind mixing, and orbital effects through the Coriolis force, all likely to produce additional variability. How these factors interact with a clumpy stellar wind is, however, beyond the scope of this work.

\subsubsection{Clump effects on the CD}\label{dynamics}

\begin{figure}
\centering
     \centering\resizebox{0.9\hsize}{!}{\includegraphics{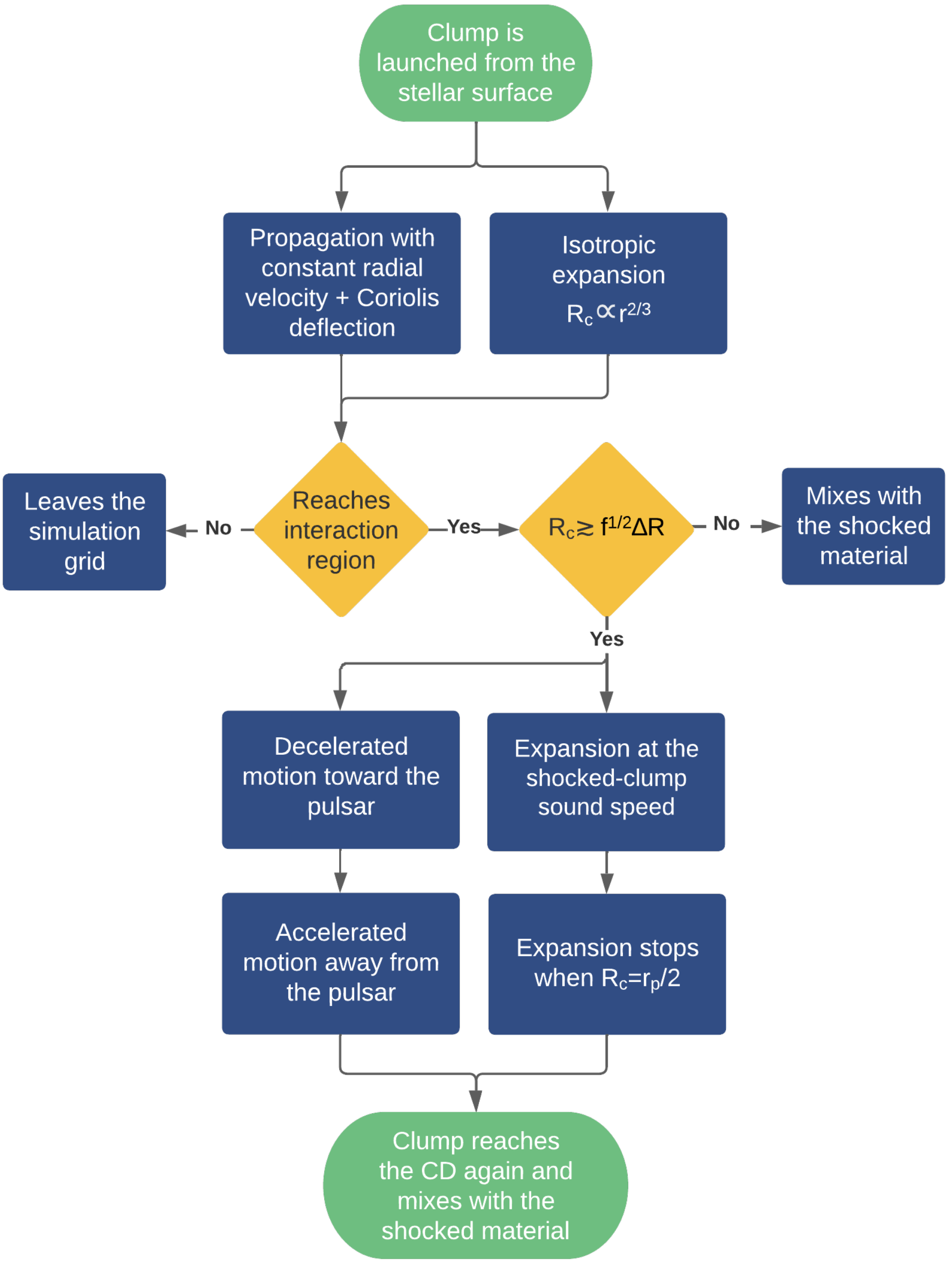}}
     \caption{Flowchart of the clump dynamics computational process.}
     \label{flowchart}
\end{figure}

To account for the nonlinear nature of the formation of clumps, we characterize their mass distribution through a power-law function: $dN_{\rm c} (M_{\rm c})/dM_{\rm c}\propto M_{\rm c}^{-k}$. The index $k$ and the volume filling factor $f$, which is the average-to-clump density ratio and links the clump size and mass, are treated as free parameters. The initial minimum and maximum clump radii adopted here are $R_{\rm c,min}=0.01\, R_{\rm{*}}$ and $R_{\rm c,max}=0.1\, R_{\rm{*}}$, respectively \citep[see Sect.~3.3 in][for a discussion]{Bosch-Ramon2013a}. Table~\ref{table:clump} summarizes the parameters selected for the four simulated cases with different degrees of clumpiness of the stellar wind. We note that $k=2$ yields a top-heavy clump distribution, whereas $k=3$ yields a bottom-heavy one. The computational method for the clump dynamics is detailed below and presented in Fig.~\ref{flowchart}.

\begin{table}
\caption{Degrees of inhomogeneity of the stellar wind}
\label{table:clump}
\centering 
\begin{tabular}{c c c}              
\hline\hline                        
\diagbox{$f$}{$k$} & 2 & 3 \\
\hline                              
0.01  & \begin{tabular}[c]{@{}c@{}}top heavy\\ high density\end{tabular} & \begin{tabular}[c]{@{}c@{}}bottom heavy\\ high density\end{tabular} \vspace{2pt}\\
0.1 & \begin{tabular}[c]{@{}c@{}}top heavy\\ low density\end{tabular}  & \begin{tabular}[c]{@{}c@{}}bottom heavy\\ low density\end{tabular} \\
\hline                              
\end{tabular}
\end{table}

Clumps, assumed to be spherically symmetric and of constant density, are isotropically launched from the stellar surface with their direction, initial mass and radius, and injection time being determined by a random number generator. The clump injection rate is chosen according to $\dot{M}$. Clumps move away from the stellar surface with a radial clump-velocity component of $u_{\rm w}$. As the calculation is done in the rotating frame, we add an azimuthal clump velocity component of the form $-r\omega_{\rm{orb}}$ (against orbital motion), where $r$ is the distance from the star, to account for the effect of the orbit. Stellar rotation is neglected at this stage. During this time, clumps are assumed to expand such that $f$ is constant, so their radii evolve as: $R_{\rm{c}}(r)=R_{\rm{c}}(R_{*})(r/R_{*})^{2/3}$.

Depending on their orientation, some clumps reach the CD, while others leave the simulation grid without interacting with the shocked two-wind region. Clumps that reach the CD may or may not be able to cross the entire shocked two-wind region, depending on their radii. The critical radius for penetration is \citep{Bosch-Ramon2013a}
    \begin{equation}
    R_{\rm{c}} \gtrsim f^{1/2} \Delta R,
    \label{mrc}
    \end{equation}
    where $\Delta R$ is the thickness of the shocked two-wind region at each location. Hydrodynamic simulations for a smooth stellar wind and $\eta$ values similar to ours \citep{Bogovalov2008,Bosch-Ramon2015} indicate that $\Delta R$ is $\sim 1/3$ of the distance to the pulsar ($r_{\rm p}$) there (which in turn depends on $\vartheta$, $\vartheta_{1}$, $\eta$, and $d$). Clumps with sizes smaller than that given in Eq.~\ref{mrc} get destroyed and dissolve in the shocked two-wind medium, not affecting the CD geometry substantially. Clumps with larger sizes can penetrate into the (until then) unshocked pulsar wind.

Once the clumps reach the pulsar-side boundary of the shocked two-wind region (i.e., the smooth pulsar wind termination shock), the ram pressure of the pulsar wind starts to impact the clumps. Adapting a semi-analytical hydrodynamical model from \cite{Barkov2012} for the evolution of a clump under the impact of a relativistic radial flow, we solve the equation of motion of the clumps in the pulsar wind zone. The radial component (outward from the pulsar) of the clump Lorentz factor evolves according to:
    \begin{equation}
    \label{eqn:motion}
    \frac{\rm{d}\Gamma_{\rm{c}}}{\rm{d}t}=\left ( \frac{1}{\Gamma_{\rm{c}}^{2}} - \frac{\Gamma_{\rm{c}}^{2}}{\Gamma_{\rm{w}}^{4}} \right ) \frac{\pi R_{\rm{c}}^{2}}{4 M_{\rm{c}}c^{2}}P_{\rm{w}}\,,
    \end{equation}
    where $P_{\rm{w}}=L_{\rm{sd}}/4\pi c r_{\rm{p}}^{2}$ is the pulsar wind ram pressure, $\Gamma_{\rm{c}}$ is the clump Lorentz factor, and $M_{\rm{c}}$ is the clump mass. The motion of the clumps is initially decelerated toward the pulsar and then, after the pulsar wind ram pressure overpowers that of the stellar wind, accelerated away from the pulsar. Gravitational forces are not dynamically relevant compared with the wind and clump thrusts in the region of interest and can be neglected.

\begin{figure*}[!ht]
\centering
\centering\resizebox{1\hsize}{!}{\includegraphics{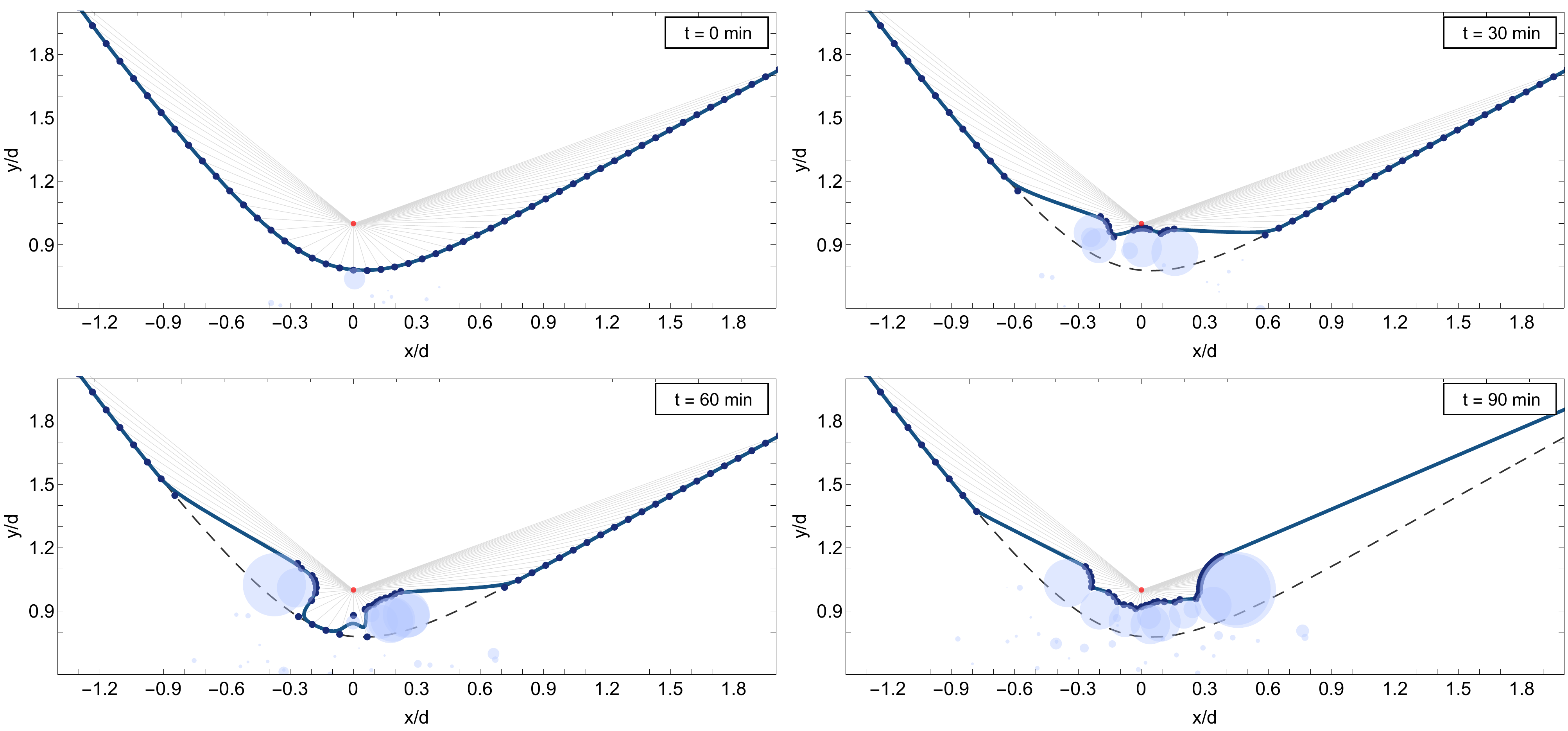}}
  \caption{Two-dimensional snapshots of a CD section at 30 minute intervals showing the derivation of the distorted CD shape. Radial lines (light gray lines) are traced from the pulsar (red point; not in scale) in all directions. If a line intersects the surface of a shocked clump (light blue disks) before reaching the smooth-wind CD (black dashed line), then a distorted-CD point is placed at this location; otherwise, the intersection point of the radial line with the smooth-wind CD is added to the grid. The new grid of points (dark blue points on the blue line) representing the distorted CD is fitted by a spline (blue solid line).}
  \label{cdder}
\end{figure*}

Within the shocked two-wind region, clumps cannot expand as they are confined by the thermal pressure of the shocked winds, but in the (previously) unshocked pulsar wind zone quick clump expansion can occur at the shocked-clump sound speed in all directions except against the pulsar wind \citep[see, e.g.,][]{delPalacio2019}. For simplicity, at this stage we assume that the shocked expanding clumps are still uniform and spherical \citep[valid at least for the initial stages of clump expansion;][]{Paredes-Fortuny2015}, with sound speed:
    \begin{equation}
    c_{\rm{s}}=\sqrt{\hat{\gamma}P_{\rm{w}}/h\rho_{\rm{c}}}\,,
    \end{equation}
    where $h=1+\hat{\gamma}P_{\rm{w}}/[(\hat{\gamma}-1)\rho_{\rm{c}}c^{2}]$ is the internal specific enthalpy, and $\rho_{\rm{c}}$ is the clump mass density. As clumps do not have time to become relativistic, the adiabatic index $\hat{\gamma}$ can be fixed to $5/3$, and both $\Gamma_{\rm c}$ and $h$ are always barely above 1.

Expansion is switched off if the radius of an expanding clump reaches half the distance to the pulsar when the undisturbed CD is initially crossed (measured from the clump center). This accounts for the fact that the approximation used for the dynamical evolution of the clumps that are shocked by the pulsar wind (computed taking the clump centers as their reference positions) is valid until an expanding clump affects much of the originally unshocked pulsar wind zone (equivalently, the clump size becomes significant with respect to $r_{\rm p}$).

At that stage, the clumps quickly get deflected away from the pulsar (as $\dot{\Gamma_{c}} \propto R_{c}^{2} $) and re-approach the original CD. Once the center of an expanded clump has intercepted that surface again, we assume that it quickly mixes with the shocked two-wind region.

We note that other $R_{\rm c}(r)$ prescriptions may be also possible because the actual clump geometry and evolution are very complex \citep[e.g.,][]{Sundqvist2018,Elmellah2020}. In the case of linear growth, for instance, the relevant quantities at the CD would be within a factor of 2 with respect to those in the adopted case, which does not qualitatively affect the conclusions of this work. The actual geometry of clumps when they fully cross the shocked two-wind region is not important, as their later expansion once shocked by the pulsar wind is fast in all directions except toward the pulsar; thus it can be considered to be roughly isotropic.

Using the above approach, which albeit crude attempts to capture the main clump evolution features found by \cite{Paredes-Fortuny2015}, we can determine the location and size of the penetrating clumps at any given time, which allows us to characterize the corresponding shape and temporal evolution of the entire distorted CD. This is done by tracing radial lines from the pulsar in all directions. If one of these lines intersects the surface of a shocked clump before the line reaches the smooth-wind CD, then a distorted-CD point is placed at the intersection location. If the line reaches the smooth-wind CD, then the CD is not distorted at this location. When this is done for all relevant directions from the pulsar, one obtains a new grid of points representing the distorted CD (see Fig.~\ref{cdder} for a sketch of the computation process). Complementary to Fig.~\ref{sketch}, Fig.~\ref{cdder} illustrates the CD between the pulsar wind and the stellar wind once it is affected by the clump presence. Clump dynamics is assumed to be affected only by the pulsar wind and not by the presence of other clumps, as the latter would be a higher-order effect and is neglected at this stage.

\subsubsection{Emitter characterization}

For simplicity, we take the CD surface as the location of the emission and divide this surface into $N$ surface elements that are treated as point-like emitters. The magnetic field and particle distribution, which are determined by the local conditions on each specific point on the CD, are treated as uniform within each individual emitter. For convenience, we express the magnetic field ($B$) energy density for each emitter as a fraction $\eta_{\rm{B}}$ of the shocked pulsar wind pressure at the CD: $u_{\rm{B}}=\eta_{\rm{B}}P_{\rm W}$. The fraction $\eta_{\rm{B}}$ is fixed throughout this work to 0.1, which corresponds to a magnetic-to-stellar photon energy density ratio of $\sim 10^{-3}$ at the CD close to the pulsar location. This $\eta_{\rm{B}}$ value lies between the weakly and strongly magnetized pulsar wind cases and is consistent with a hydrodynamics approach to characterize the CD. Taking different $\eta_{\rm{B}}$ values would not significantly change our conclusions, as radiation losses would still be dominated by IC and because our main focus is on the relative flux change in specific energy bands (although broadband spectral changes would be expected). Only electron/positron cooling processes are considered at this stage, that is, IC and synchrotron emission, because they are the most efficient radiation processes in compact binaries \citep[see, e.g.,][]{Bosch-Ramon2009}.

The complexity of the emitter structure is very high, as shown by previous works dealing only with some of the aspects of the scenario studied here \citep[see, e.g.,][where RHD simulations were combined with radiative calculations]{Dubus2015,delaCita2017a,Huber2020a}. However, as we are interested in the typical clump-induced emission time variability at certain energies and not in deriving a detailed SED, we simplify the $N$ point-like emitters as being in one of two extreme cooling regimes: radiative or adiabatic.

\begin{figure*}[!ht]
  \centering\resizebox{1\hsize}{!}{\includegraphics{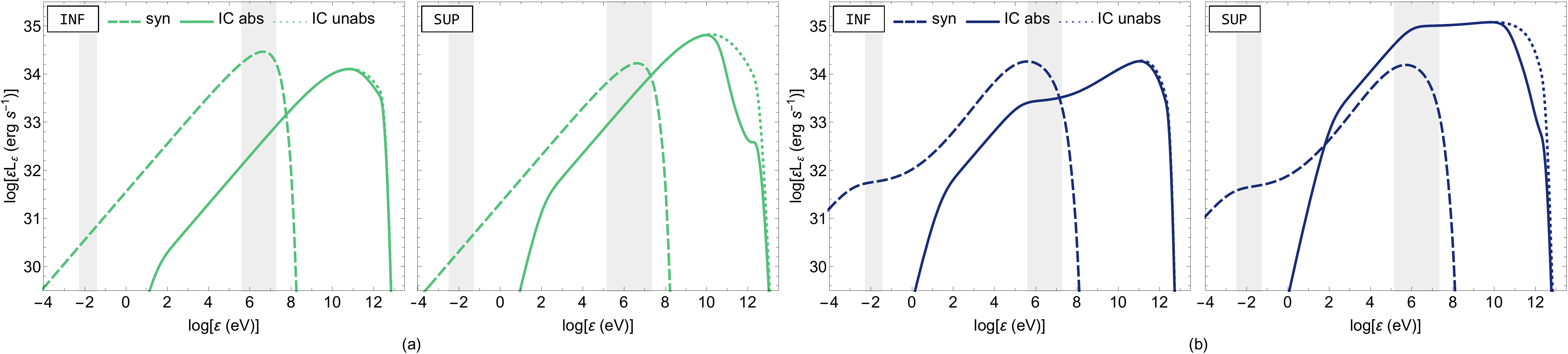}}
     \caption{SEDs for the smooth-wind case for the (a) adiabatic (green) and (b) radiative (blue) regimes considering regions up to $1.5d$ from the star at inferior conjunction (INF; $\phi=0.716$) and superior conjunction (SUP; $\phi=0.058$). The synchrotron (colored dashed) and IC (colored solid) components are plotted, and the unabsorbed spectra (colored dotted) are also shown for comparison. As a reference, the shaded areas indicate the synchrotron ($\sim 10^{-2}$~eV) and IC ($\sim 10^{7}$~eV) photon energies corresponding to the (adiabatic-to-radiative) transition particle energy for all individual emitters in the CD.}
     \label{SED}
\end{figure*}

In the fully radiative regime, the energy of the emitting particles is radiated away right after reaching the CD. In this case, the particle energy distribution is determined by the acceleration (injection) energy distribution and the relevant radiative losses only (see below). In the adiabatic regime, the total energy density in the post-shock region, $\approx 3\,P_{\rm w}$ times the relevant volume (known only approximately; see below), determines the nonthermal population there. In this case, the particle energy distribution coincides with the acceleration one. Intermediate cases are more realistic, with an expected transition between regimes from the higher particle energies (radiative; more important close to the pulsar) to the lower particle energies (adiabatic; more important far from the pulsar). This transition would manifest itself in features of the IC and synchrotron SED that are hinted when comparing our results in both regimes. Albeit simple, our approach allows for the derivation of general conclusions on the radiation behavior because results for both cooling regimes are provided. The normalizations of the particle distributions in both regimes are just approximately consistent with each other due to our simplified treatment of the hydrodynamics in the adiabatic case.

In our model, relativistic particles are injected following a power-law distribution in energy, $E$, with an exponential cutoff of the form
\begin{equation}
\label{eqn:eldis}
\Delta Q(E) \propto E^{-p} \exp[-E/E_{\rm{max}}],
\end{equation}
where we adopt $p=2$ as a fiducial value. 
Significantly softer (harder) injection energy distributions would yield substantially lower (higher) fluxes in the higher-energy regions of the synchrotron and IC components. Further, injection energy distributions more complex than a power-law function, such as an additional narrow component peaking around $\Gamma_{\rm w}$ \citep[see, e.g.,][]{Dubus2015}, would lead to more complex SEDs. Again, given our focus on the clump-induced relative variability within specific energy bands, the adopted injection distribution is enough for our purposes.

The maximum electron energy, characterized by $E_{\rm{max}}$ in Eq.~\ref{eqn:eldis}, is mainly constrained by acceleration, radiative losses, and escape. As particles with multi-TeV energies exist in gamma-ray binaries, we assume a sufficient acceleration rate, with efficiency $\eta_{\rm{acc}}=0.1$ and timescale $t_{\rm{acc}}=E/(\eta_{\rm{acc}}qcB)$, where $q$ is the electron charge. As noted, the relevant cooling channels in the shocked pulsar wind are synchrotron emission and IC. The energy-loss rate for synchrotron ($\dot{E}_{\rm{syn}}$) is derived from \cite{Blumenthal1970} and for IC ($\dot{E}_{\rm{IC}}$) from \cite{Khangulyan2014} assuming an anisotropic distribution of blackbody photons. Equating the acceleration timescale $t_{\rm{acc}}$ with the total radiation cooling timescale,  $t_{\rm{cool}}=E/(|\dot{E}_{\rm{syn}}|+|\dot{E}_{\rm{IC}}|)$, yields $E_{\rm{max}}^{\rm{cool}}$. We also consider the Hillas criterion \citep{Hillas1984}, which is similar to the constraint from advection escape and gives an upper limit of the maximum energy: $E_{\rm{max}}^{\rm{Hillas}}=qBr_{\rm{s}}$, where the spatial scale $r_{\rm{s}}$ represents the size of the source, taken here simply as $\sim d$. Then, $E_{\rm{max}}=\rm{min}$ $(E_{\rm{max}}^{\rm{cool}},E_{\rm{max}}^{\rm{Hillas}})$, which is typically $\sim 10$~TeV.

The minimum electron distribution energy has been fixed to $E_{\rm{min}}=100$~MeV. This parameter strongly affects the lowest-energy part of the IC SED and weakly the normalization of the distribution of particles when their total energy is fixed. The former affects the contribution of IC emission below X-rays, whereas the latter effect can be ``absorbed'' by changing the total energy in the power-law distribution of particles. This nonthermal energy, which affects the normalization of the SED linearly, is taken here to be the entirety of the shocked pulsar wind energy, but it can be lower (e.g., most of the energy may be in a narrow particle energy distribution). 



In the radiative regime, the electron injection distribution is normalized by the pulsar wind luminosity within the solid angle of each emitter, $\Delta L_{\rm{\Omega}}$:
\begin{equation}
\int_{E_{\rm{min}}}^{E_{\rm{max}}}\Delta Q(E)E\textup{d}E=\Delta L_{\rm{\Omega}}.
\end{equation}
The corresponding cooled particle population can be described as:
\begin{equation}
\Delta N(E)=|\dot{E}|^{-1}\int_E^{E_{\rm max}} \Delta Q(E'){\rm d}E'.
\end{equation}
On the other hand, in the adiabatic regime the electron distribution $\Delta N(E)$ follows the energy dependence in Eq.~\ref{eqn:eldis} and is normalized by the internal energy $\Delta E$ stored within an emitter of approximate volume $\textup{d}V=\textup{d}S_{\textup{sph}} h_{\rm{s}}$, where $\textup{d}S_{\textup{sph}}=r_{\textup{p}}\textup{d}\Omega$, and $h_{\rm{s}}$ is the thickness of the shocked pulsar wind shell. To determine $h_{\rm{s}}$, we take $h_{\rm{s}}=0.2\,r_{\rm p}$ based on relativistic hydrodynamical simulations \citep[e.g.,][]{Bosch-Ramon2015}. Thus, one has:
\begin{equation}
\int_{E_{\rm{min}}}^{E_{\rm{max}}}\Delta N(E)E\textup{d}E= \Delta E.
\end{equation}

Once the particle energy distribution for each point-like emitter has been obtained, the associated synchrotron and IC SEDs are computed using the formulae given in \cite{Blumenthal1970} and \cite{Khangulyan2014}, respectively. Given that the flow is only weakly relativistic in the subsonic region of the shocked pulsar wind and the orientation distribution of the flow velocity is broad in the CD (even broader in the case of a distorted CD; \citealt{Paredes-Fortuny2015}), Doppler boosting has not been taken into account. Gamma-ray absorption in the stellar photon field through pair creation has been calculated using the cross section from \cite{Gould1967}. The radiation of the created pairs may affect the X-ray and gamma-ray emission \citep[e.g.,][]{Bednarek2007,Bosch-Ramon2008}, but we neglect this contribution at this stage.  
Due to the clump presence and orbital motion, changes in the interaction angles of IC and gamma-ray absorption and in the CD geometry (also affecting synchrotron radiation through $B$) must be taken into account.


\section{Results}\label{res}

\subsection{Smooth case}\label{sec:smooth}

We have computed the SEDs and light curves in the case of a smooth stellar wind along the entire orbit including regions of the CD up to $1.5d$ from the star; cases with different emitting sizes mostly affect the normalization and are not presented here. Figure~\ref{SED} shows the synchrotron and IC SEDs for the two cooling regimes (adiabatic and radiative) and two orbital phases: inferior ($\phi=0.716$ -INF-) and superior conjunction ($\phi=0.058$ -SUP-).

The results obtained are similar to those found in previous works (see Sect.~\ref{intro}). In the adiabatic regime, synchrotron and IC radiation are rather hard, with the synchrotron component peaking at soft gamma rays, and the IC component softening above $\sim 1-10$~GeV due to the Klein--Nishina (KN) effect in the cross section and peaking around $\sim 10-100$~GeV. In the radiative regime, the synchrotron peak broadens, but the SED is still hard down to $\sim 10$~eV due to IC losses in the KN limit, where it becomes flat due to dominant IC losses in the Thomson regime. The (unabsorbed) IC SED is still moderately hard at $\sim 0.1-100$~GeV due to IC KN losses, softens above that energy due to synchrotron becoming dominant plus the KN effect, and flattens below $\sim 0.1$~GeV due to IC Thomson losses. Below $\sim 1$~MeV, corresponding to (cooled) electrons with $E<E_{\rm min}$, the IC spectrum hardens again. The effect of gamma-ray absorption in the SED is the strongest and starts at the lowest gamma-ray energy ($\sim 30$~GeV) at superior conjunction when the photon--photon interaction angle is the largest. Absorption becomes less severe the farther the pulsar is from that orbital phase, becoming a minor effect around inferior conjunction, mostly due to the interaction angle decreasing \citep[an effect smoothed further by the extended emitter; e.g.,][]{Khangulyan2008}. Pulsar--star distance changes are a secondary factor for small-to-moderate eccentricities that enhances absorption toward periastron. For simplicity, soft X-ray absorption due to the photoelectric effect has not been considered at this stage. We did not study radio wavelengths either, as this emission is expected to be severely free--free absorbed so close from the star \citep[see, e.g.,][and references therein]{Molina2020}. 

\begin{figure}[!ht]
  \centering\resizebox{1\hsize}{!}{\includegraphics{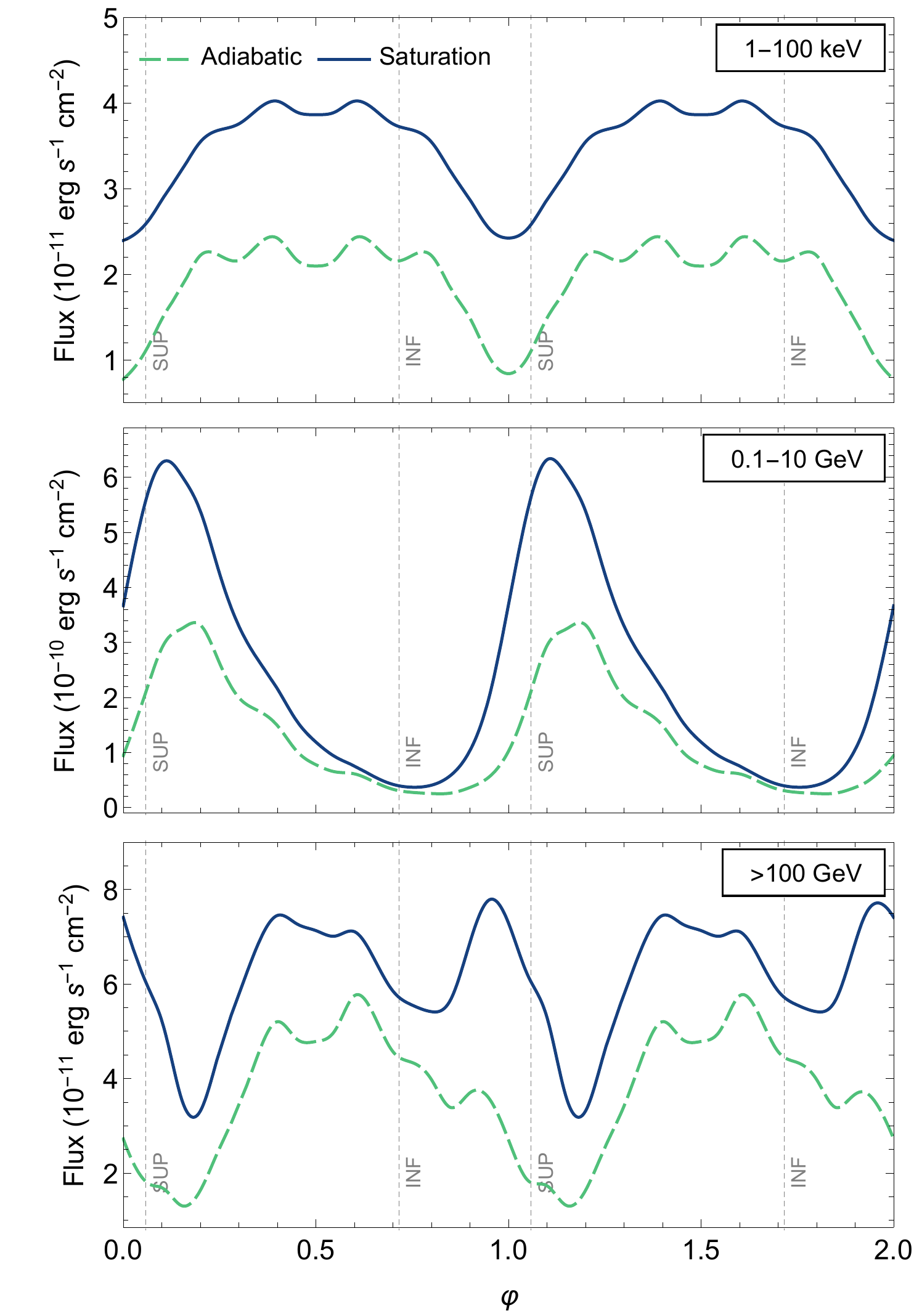}}
  \caption{Light curves for a smooth stellar wind for the adiabatic (green dashed) and radiative (blue solid) regimes including regions of the CD up to $1.5d$ from the star for synchrotron emission in the 1--100 keV band (top) and IC emission in the 0.1--10 GeV (middle) and >100 GeV (bottom) energy bands. The vertical dashed lines labeled SUP and INF mark the superior and inferior conjunction, respectively. We show two orbital cycles for clarity.}
  \label{lc}
\end{figure}

Figure~\ref{lc} shows the orbital light curves in the case of a smooth stellar wind for the radiative and adiabatic regimes and three energy bands: X-rays (1--100~keV; synchrotron), HE gamma rays (0.1--10 GeV; IC), and VHE gamma rays ($>100$~GeV; IC). For the X-ray light curves, we only show the synchrotron components (adiabatic and radiative), because IC X-rays are produced deep in the adiabatic regime (see Fig.~\ref{SED}), and the corresponding curve is lower than any synchrotron curve. The HE and VHE gamma-ray light curves are by contrast shown only for IC (both regimes), which is the dominant emission at those energies. The light curves are mostly explained by changes in the emitter spatial volume and angular size as seen from the pulsar (affecting particle normalization), in the distance from the emitter to the pulsar (affecting $B$ and synchrotron in the X-rays) and the star (more mildly affecting IC target photons), and by angular effects (affecting IC in the HE and VHE and gamma-ray absorption in the VHE). The fact that the emitter is extended with size $\sim d$ and has a complex geometry changing due to eccentricity introduces some variations, but overall the results are as expected from previous work, with X-rays peaking around apastron, and the HE and the VHE gamma-rays peaking around superior and inferior conjunction, respectively. The VHE light curve presents a secondary peak before periastron, but similar features were already found in previous work \citep[e.g.,][]{Khangulyan2008,Dubus2008,Takahashi2009,Molina2020}. 

\subsection{Clumpy-wind case}\label{sec:clumpy}

\begin{figure}
  \centering\resizebox{1\hsize}{!}{\includegraphics{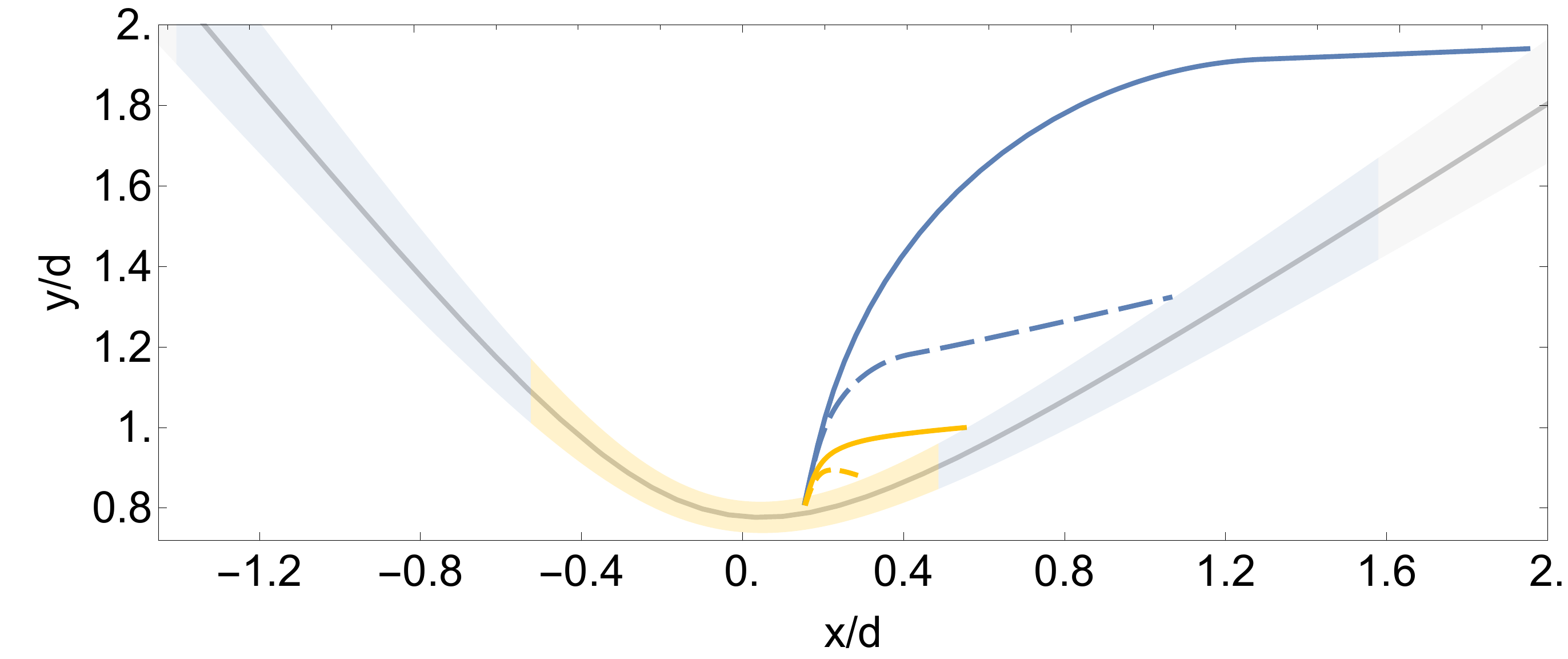}}
  \caption{Two-dimensional cut on the orbital plane of the smooth CD (solid gray line) at an intermediate orbital phase ($\phi=0.28$) showing the trajectories of multiple clumps. The light-gray shaded area gives a rough estimate of the thickness of the shocked two-wind region. The blue and yellow lines correspond to high-density and low-density clumps for stellar wind filling factors of $f=0.01$ and $f=0.1$, respectively. The clump radii when launched from the stellar surface are $0.1 R_{*}$ (solid) and $0.06 R_{*}$ (dashed). The colored shaded areas show the regions along the pulsar wind shock front within which the largest high-density (blue shaded area) and low-density (yellow shaded area) clumps can penetrate.}
  \label{cltra}
\end{figure}

To study the short-term variability induced by a clumpy stellar wind, we focus on an intermediate orbital phase ($\phi=0.28$), at which the line of sight is perpendicular to the pulsar--star line, and the three mentioned energy bands (X-rays and HE and VHE gamma rays) and cooling regimes (adiabatic and radiative). In Fig.~\ref{cltra}, we show the geometry on the orbital plane of the smooth CD and the region filled with the shocked two-wind flow. We also show the trajectories of clumps of different initial masses/radii and densities that manage to penetrate the shocked two-wind region close to the stagnation point, where the minimum initial clump radii for penetration are $\sim 0.02 R_{*}$ for $f=0.01$ and $\sim 0.05R_{*}$ for $f=0.1$. 

\begin{figure}
  \centering\resizebox{0.9\hsize}{!}{\includegraphics{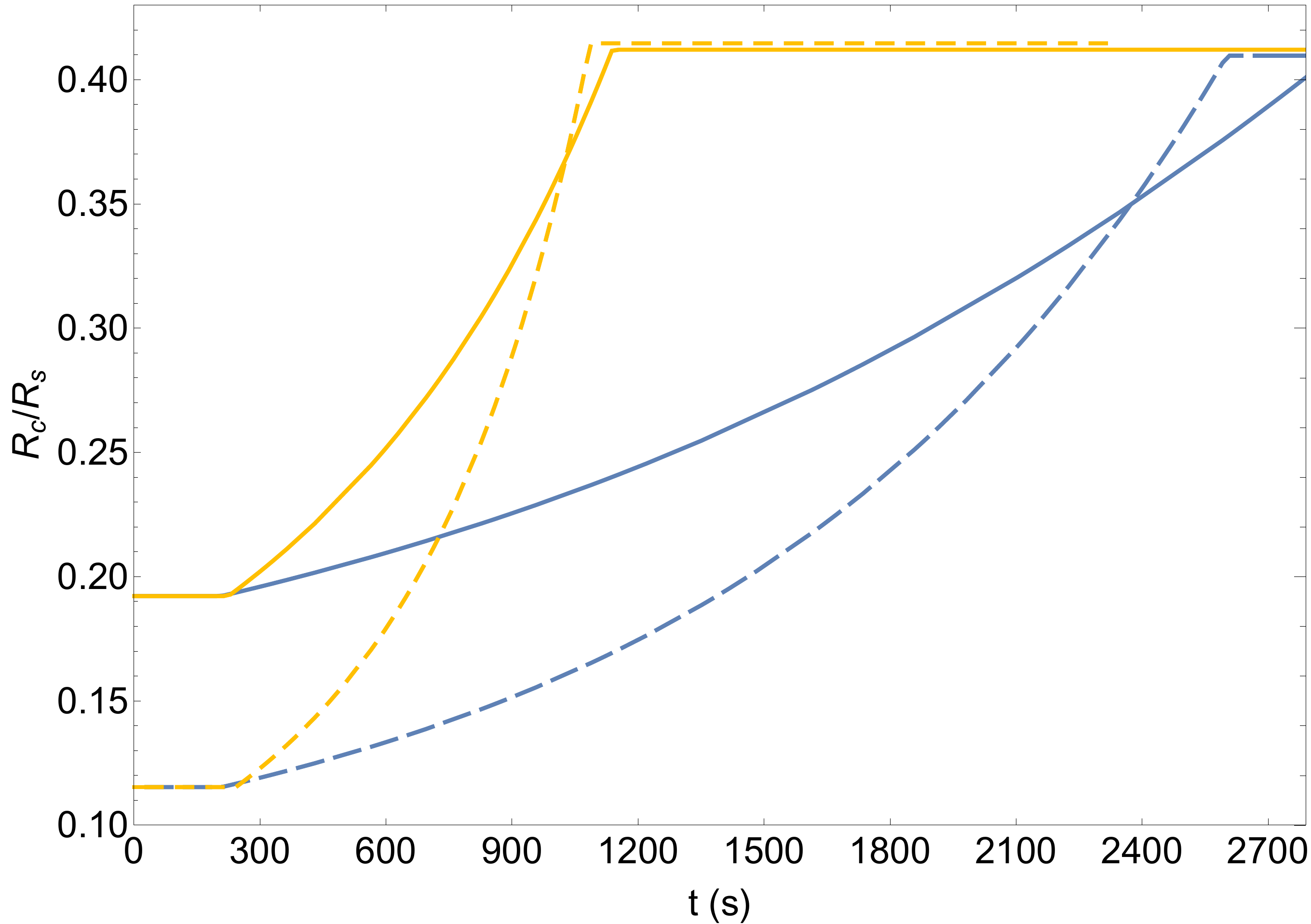}}
  \caption{Evolution over time of the radii of the clumps whose trajectories are shown in Fig.~\ref{cltra}. Line patterns are the same as in Fig.~\ref{cltra}.}
  \label{clevol}
\end{figure}

Although the largest clumps with $f=0.01$ can reach the unshocked pulsar wind region almost everywhere in the explored CD region (blue shaded area in Fig.~\ref{cltra}), for the largest clumps with $f=0.1$ the CD crossing is restricted to the vicinity of the stagnation point (yellow shaded area). Figure~\ref{clevol} shows the time evolution of the radii of these clumps from the time of entering the shocked two-wind region until the time clump expansion is halted. All these clumps reach the same final size (half $r_{\rm p}$ at the crossing point) because they share the same penetration location; although, smaller or lighter clumps tend to evolve faster. Clumps entering farther away from the smooth-CD symmetry axis may not reach again the smooth-CD location within the simulated region.

\begin{figure*}[!ht]
 \centering\resizebox{1\hsize}{!}{\includegraphics{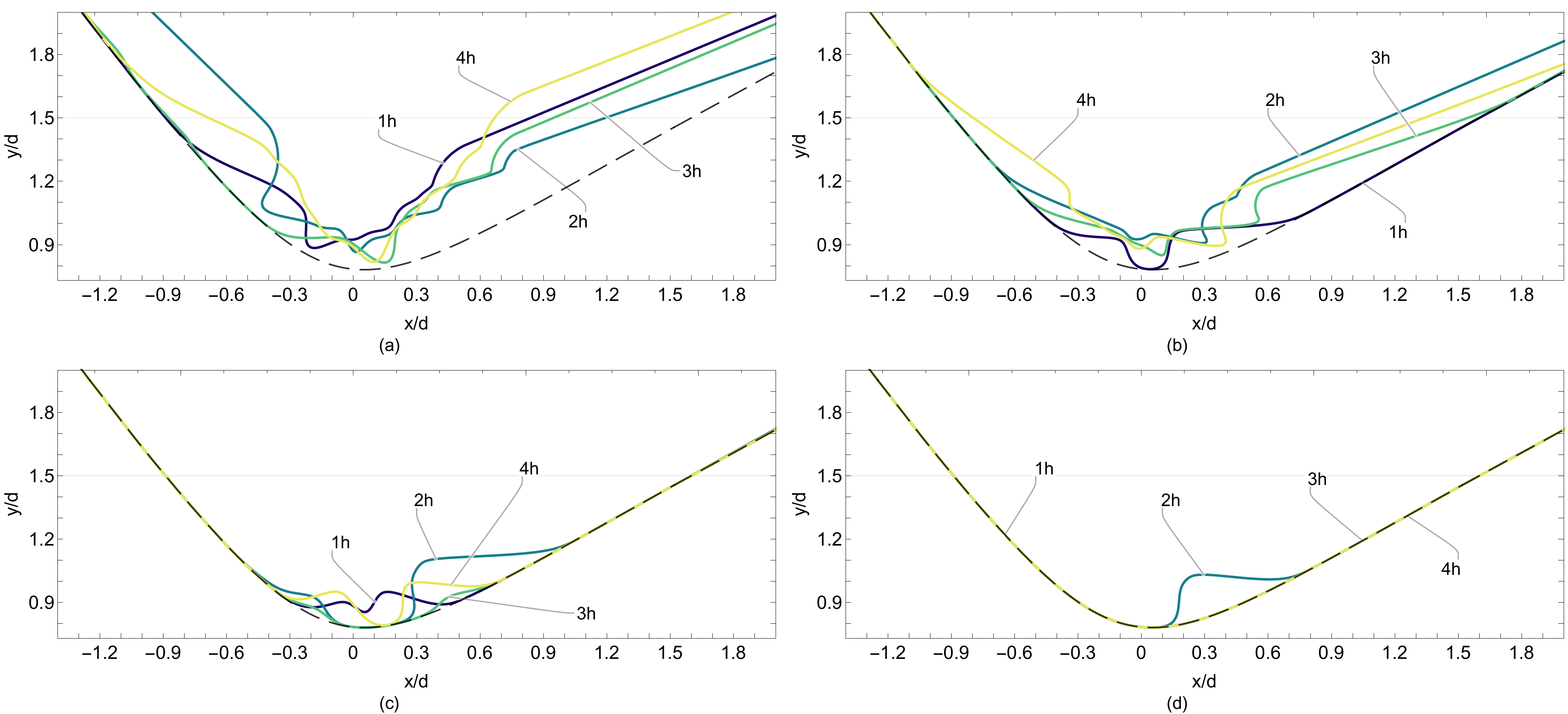}}
\caption{Two-dimensional snapshots of the CD on the orbital plane at 1 hour intervals within a period of 3 hours for (a) a top-heavy ($k=2$) high-density ($f=0.01$), (b) a bottom-heavy ($k=3$) high-density ($f=0.01$), (c) a top-heavy ($k=2$) low-density ($f=0.1$), and (d) a bottom-heavy ($k=3$) low-density ($f=0.1$) clump distribution. The black dashed lines correspond to a smooth stellar wind. The horizontal gray lines mark the CD size used to compute the light curves in Fig.~\ref{cllc}.}
\label{clCD}
\end{figure*}

In Fig.~\ref{clCD}, we plot cuts of the shocked two-wind region on the orbital plane, for stellar winds of different degrees of inhomogeneity over a period of 3 hours at 1 hour intervals. The arrival of clumps to the CD deforms the overall interaction structure, generating relatively quick and global variations in the geometry of the emitting shocked pulsar wind. 
In the most extreme case (top-left panel), clumps frequently reach quite close to the pulsar, radically distorting the CD geometry. Otherwise, for a bottom-heavy low-density clump distribution (bottom-right panel), the results in the explored narrow phase interval essentially coincide with a smooth-wind scenario mainly because of the small number of clumps that are large enough to cross the shocked two-wind region. In intermediate cases (bottom-left and top-right panels), moderate but significant CD variations may still be seen.

\begin{figure}[!ht]
  \centering\resizebox{1\hsize}{!}{\includegraphics{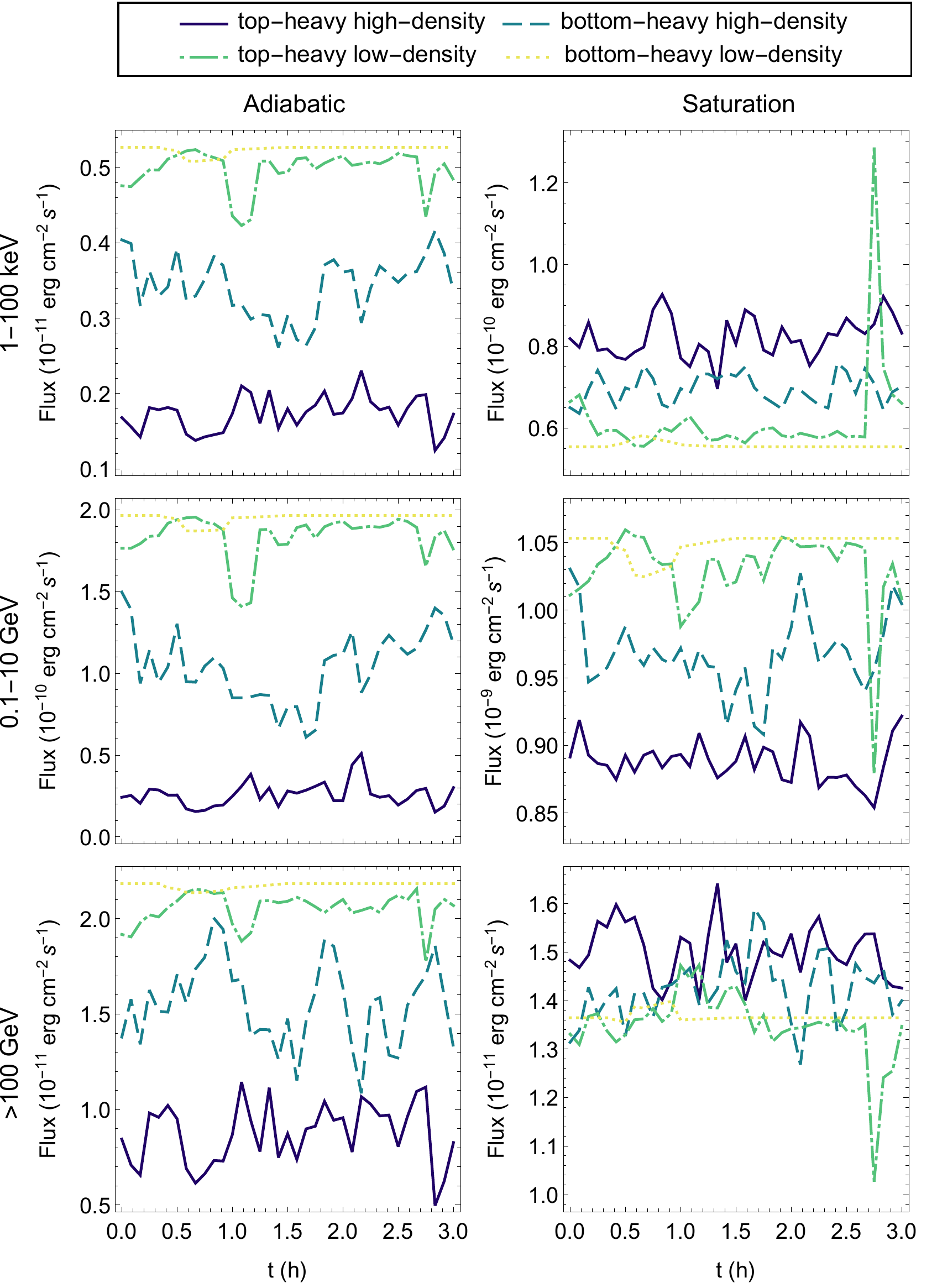}}
  \caption{Light curves over a period of 3 hours for the adiabatic (left column) and radiative (right column) regimes at energies 1--100 keV (synchrotron only; top rows) and 0.1--10 GeV and >100 GeV (IC only; middle and bottom rows, respectively) for a top-heavy high-density (($k=2$, $f=0.01$; solid), a bottom-heavy high-density ($k=3$, $f=0.01$; dashed), a top-heavy low-density ($k=2$, $f=0.1$; dashed--dotted), and a bottom-heavy low-density ($k=3$, $f=0.1$) clump distribution (dotted).}
  \label{cllc}
\end{figure}

Figure~\ref{cllc} shows the light curves for different stellar wind scenarios in both cooling regimes over a period of 3 hours. The light curves are computed considering an emitter size extending up to a distance of $1.5\,d$ from the star. Referring to the adiabatic X-ray light curves (top left panel) as an example, one can see that the light curves of the two most inhomogeneous stellar winds (blue solid and dashed lines) exhibit large and rapid variability. For a relatively low degree of clumpiness (green dashed--dotted line), large flaring events still occur, albeit more sparsely, while for the least clumpy case (yellow dotted line) variability subsides almost completely. The same trends in variability are observed in all panels. The light curves for the smooth-wind case are not shown for simplicity but largely coincide with the bottom-heavy low-density light curves.

Using Fig.~\ref{SED}, one can evaluate which light curve, either radiative or adiabatic, is more realistic. In all the studied energy bands, the emission in the region of interest is produced in the radiative regime, and thus, despite the simplicity of the approach, this regime provides to first order a reasonably realistic account of the behavior of the emission. In this regime, the relative changes in flux are moderate in the explored narrow phase interval, of $\sim 20$\% (top-heavy and bottom-heavy dense clumps), 10--100\% (top-heavy light clumps; note the large X-ray spike in the top right panel), and $\lesssim 10$\% (bottom-heavy light clumps), with variability timescales of $\sim 0.1-1$~h.

It is worth noting that the adiabatic light curves present significantly higher variability than the radiative ones in the dense-clump cases. Thus, although radiative losses are generally dominant on the scales we explored, including supersonic flow regions (not considered here) could lead to higher variability, as adiabatic losses can become stronger there \citep{Khangulyan2014b}. For the emission produced more deeply in the radiative regime, the impact of hydrodynamics is expected to be small.

We also assessed whether there is a reduction in variability due to emission coming from CD regions potentially more or less affected by clumps. We compared our results with light curves for emitting regions extending closer (up to $1.2\,d$) or farther from the star (up to $1.8\,d$). Larger sizes of the emitting region do not smooth out the relative changes in the flux, mostly affecting just the normalization, which indicates that the clump presence still affects the emitter relatively far from the pulsar. 


\section{Discussion}\label{disc}

In general, the predicted luminosities in the explored energy bands are loosely similar to those of LS~5039 at a few-kpc distance. For the corresponding fluxes, present instrumentation may be able to trace the short-term relative variations in X-rays and VHE gamma rays \citep[see typical rate uncertainties in, e.g.,][for X-rays and VHE gamma rays, respectively]{Martocchia2005,Aharonian2006}, and the future Cherenkov Telescope Array would reach significantly finer levels of variability at VHE \citep[e.g.,][]{Paredes2013}.
At HE gamma rays, several orbital cycles must be folded in phase to get statistically meaningful orbital light curves, which prevents studying (nonextreme) variability on suborbital timescales.

\emph{Clump interaction rate:} The frequency of interaction of the largest clumps with the shocked pulsar wind can be estimated from the mass distribution as:
\begin{equation}
\dot{N}\sim M_{\rm c,max} \frac{dN(M_{\rm{ c,max}})}{dM_{\rm c}} \frac{\Delta \Omega}{4\pi},
\end{equation}
where the solid angle of interaction $\Delta \Omega = \pi (d_{\rm p}/d)^2$ depends on the radius $d_{\rm p}$ of the region that can be penetrated by clumps. For a moderately clumpy stellar wind (i.e., top heavy, low density), $d_{\rm p}\approx 0.5d$ (yellow shaded area in Fig.~\ref{clevol}), which yields an interaction rate of $\dot{N}\sim6\times 10^{-5}$~s$^{-1}$ or $\sim 21$ clumps per orbit and an average time between consecutive interactions of $t_{\rm c}=\dot{N}^{-1}\sim1.6\times 10^4$~s. The duration of each individual interaction is on the order of the clump residence time in the region $\tau_{\rm c}=d_{\rm p}/u_{\infty}\sim6\times10^3$~s. This implies that approximately 40\% of the time one large clump is interacting with the CD, and $\approx 14$\% of the time two large clumps may simultaneously interact with the CD, meaning several times per orbit. Even in the least clumpy case explored here, a few of the largest clumps would still be expected to (individually) interact with the CD per orbit.

\emph{Effect of $\eta$:} Although our study focused on $\eta=0.08$, we have run a few trials in low resolution for different $\eta$ values for completeness.
The increase in the shocked two-wind shell thickness of the termination shock for larger $\eta$ values shrinks the penetrable region to the very close vicinity of the pulsar. Larger-mass clumps that can penetrate tend to grow significantly and cause large distortions of the CD, but the overall effect is a reduced variability due to a low rate of interactions. Otherwise, for smaller $\eta$ values, the termination shock wraps closer to the pulsar, extending the penetrable region to almost the entire CD length, effectively increasing the interaction rate. Clumps entering relatively close to the stagnation point will have long residence times (although expansion will halt soon after crossing), while those entering from the outer regions will evolve and exit more quickly. In this second case, overall, variability becomes faster and more intense than in the fiducial calculations.

\section{Summary}\label{summ}

We have studied the effects of different clumpy stellar winds interacting with a relativistic pulsar wind in a high-mass binary, adopting different filling factors ($f=0.1$ and 0.01) and levels of clumpiness ($dN_{\rm c}/dM_{\rm c}\propto M_{\rm c}^{-2}$, top heavy; and $\propto M_{\rm c}^{-3}$, bottom heavy). Our results for the SED and orbital light curves in the smooth-wind case are similar to those presented in the literature. The presence of clumps does not seem in general to strongly modify the emitter, and thus smooth-wind models seem enough to capture the most prominent behavior of the source, but there is clear clumpy-wind-induced variability in the X-ray and HE and VHE gamma-ray light curves that can be detectable by high-energy instrumentation.

Regarding the different cases explored, one finds that the bottom-heavy light-clump distribution can be, most of the time, indistinguishable from a smooth wind in the wind-interaction region explored in this work. In the remaining cases (denser and/or top-heavy clump distributions), the presence of clumps may be traced in the X-ray light curve and less prominently in the VHE gamma-ray light curve, with predicted flux variations of $\sim 10-20$\% in the radiative regime. These variations could reach $\sim 100$\% in some rare events, that is, the arrival to the CD of particularly large clumps or even of several large clumps simultaneously. These events, despite being more sporadic, would be very noticeable and can occur even in the least clumpy case explored in this work. Adding extra factors such as clump-induced effects further downstream of the emitting flow may reveal a stronger impact of clumps.


\begin{acknowledgements}
The authors want to thank the referee, Achim Felmeier, for constructive and useful comments that helped to improve the manuscript. The authors wish to acknowledge Edgar Molina for helpful discussions. This work has received financial support from the State Agency for Research of the Spanish Ministry of Science and Innovation under grant PID2019-105510GB-C31 and through the ``Unit of Excellence Mar\'ia de Maeztu 2020-2023'' award to the Institute of Cosmos Sciences (CEX2019-000918-M). V.B-R. is Correspondent Researcher of CONICET, Argentina, at the IAR.  
\end{acknowledgements}

\bibliographystyle{aa} 
\bibliography{ref} 

\end{document}